\documentclass[letterpaper,english,prd,preprintnumbers,nofootinbib,showpacs]{revtex4}
\usepackage[T1]{fontenc}
\usepackage[latin9]{inputenc}
\usepackage{amsmath}
\usepackage{makeidx}

\makeatletter

\usepackage{babel}

\begin{document}

\preprint{Alberta Thy 21-14}

\title{Magnetic moment of the positronium ion and virial relations}

\author{Andrzej Czarnecki and Yi Liang}

\affiliation{Department of Physics, University of Alberta, Edmonton, Alberta,
Canada T6G 2E1}
\begin{abstract}
We derive a hypervirial relation for the positronium ion, a three-body
bound state of two electrons and a positron. It relates expectation
values of three operators and reconciles two recently published, seemingly
inequivalent formulas for the magnetic moment of the ion. As a by-product,
the precision of the leading binding correction is improved. 
\end{abstract}

\maketitle
The positronium ion is no longer an exotic system. This very weakly
bound state of two electrons and a positron can now be copiously produced
with a recently discovered method \cite{PsIonSource08,PsIonSource12}:
up to two percent of positrons can be converted into ions $\mathrm{Ps}^{-}$.
This advance stimulates theoretical studies of the properties of the
ion, including its magnetic moment. The two electrons in $\mathrm{Ps}^{-}$ form
a spin singlet so the entire magnetic interaction is due to the positron,
whose gyromagnetic ratio is however slightly modified by the binding.
This binding effect, of the order of the fine structure constant squared,
$\alpha^{2}$, has recently been determined using two different approaches
\cite{Liang:2014gwa,Wienczek2014}, the latter one more precise. The
result is the expectation value of a combination of three operators,
involving the kinetic and the potential energy of the positron, and
a correlation of position vectors of the three particles. 

The two studies agree on the total numerical value, but find quite
different relative sizes of the three contributions. It has been thought
that the two expressions are not equivalent \cite{Wienczek2014},
casting doubt on the correctness of at least one of the approaches.
Here we derive an identity that relates the three expectation values
and thus reconciles both results.

The identity is of the hypervirial type
\cite{hypervirial1960,ShabaevVirial91}.  In addition to revealing the
equivalence of the two published results
\cite{Liang:2014gwa,Wienczek2014}, it allows us to eliminate one of
the three expectation values and thus present a new expression,
simpler and more precise, for the magnetic moment of $\mathrm{Ps}^{-}$.

A challenge in studying $\mathrm{Ps}^{-}$ is that the wave function of this
three-body system is not known analytically. It is computed with the
variational method, using the nonrelativistic approximation. In order
to present the Hamiltonian, we denote the coordinates of the two electrons
by $\vec{r}_{1}$ and $\vec{r}_{2}$, and use $\vec{r}_{3}$ for the
positron. The internal dynamics depends only on the relative coordinates
$\vec{r}_{12}\equiv\vec{r}_{2}-\vec{r}_{1}$ and $\vec{r}_{13}\equiv\vec{r}_{3}-\vec{r}_{1}$,
and not on the position of the center of mass $\vec{R}=\frac{1}{3}\left(\vec{r}_{1}+\vec{r}_{2}+\vec{r}_{3}\right)$.
The momenta of the three particles become\begin{eqnarray}
\vec{p}_{1} & \to & i\vec{\nabla}_{\vec{r}_{12}}+i\vec{\nabla}_{\vec{r}_{13}}\label{eq:p1}\\
\vec{p}_{2} & \to & -i\vec{\nabla}_{\vec{r}_{12}}\label{eq:p2}\\
\vec{p}_{3} & \to & -i\vec{\nabla}_{\vec{r}_{13}},\label{eq:p3}\end{eqnarray}
and the Hamiltonian, written in atomic units (and with $\vec{\nabla}_{\vec{r}_{ij}}\equiv\vec{\nabla}_{ij}$)
\begin{equation}
H=-\nabla_{12}^{2}-\nabla_{13}^{2}-\vec{\nabla}_{12}\cdot\vec{\nabla}_{13}+\frac{1}{r_{12}}-\frac{1}{r_{13}}-\frac{1}{r_{23}}.\label{eq:Ham}\end{equation}

The bound positron $g$-factor is the sum of the free particle value
$g_{\mathrm{{free}}}$ and the binding correction $\Delta g_{\mathrm{bound}}$. The three expectation
values needed to express $\Delta g_{\mathrm{bound}}$ are\begin{eqnarray}
A & = & \left\langle p_{3}^{2}\right\rangle =-\left\langle \nabla_{13}^{2}\right\rangle =0.257\,532\,962\label{eq:A}\\
B & = & \left\langle \frac{1}{r_{13}}\right\rangle =0.339\,821\,023\label{eq:B}\\
C & = & \left\langle \frac{\vec{r}_{13}\cdot\vec{r}_{23}}{r_{13}^{3}}\right\rangle =0.046\,478\,421,\label{eq:C}\end{eqnarray}
where the numerical values, taken from \cite{Wienczek2014}, are presented
here only to allow the reader to check the relation among them, derived
below. The values of $A$ and $B$ agree with the more precise evaluations
by Frolov \cite{Frolov2007}. The results for $\Delta g_{\mathrm{bound}}$ found in the
recent studies are (for now neglecting self-interaction corrections,
taking $g_{\mathrm{{free}}}\to2$) \begin{eqnarray}
\mbox{Ref. \cite{Liang:2014gwa}}\qquad\frac{\Delta g_{\mathrm{bound}}}{\alpha^{2}} & = & -\frac{11}{9}A-\frac{2}{3}\left(B-C\right),\label{eq:Liang}\\
\mbox{Ref. \cite{Wienczek2014}}\qquad\frac{\Delta g_{\mathrm{bound}}}{\alpha^{2}} & = & -A-\frac{22}{27}B+\frac{14}{27}C.\label{eq:Wienczek}\end{eqnarray}
In order to show the equivalence of these expressions, consider the
expectation value of the commutator of the Hamiltonian \eqref{eq:Ham}
with the operator $\vec{r}_{23}\cdot\vec{\nabla}_{13}$. Taken in
a stationary state, such an expectation value must vanish, since it
expresses the change with time of a time-independent operator \cite{virialAJP}.
Evaluating the commutator we find the identity\begin{equation}
0=\left\langle \left[\vec{r}_{23}\cdot\vec{\nabla}_{13},H\right]\right\rangle =\left\langle \nabla_{13}^{2}-\vec{\nabla}_{12}\cdot\vec{\nabla}_{13}+\frac{1}{r_{23}}+\frac{1}{r_{13}}-\frac{\vec{r}_{12}\cdot\vec{r}_{13}}{r_{13}^{3}}\right\rangle .\label{eq:id}\end{equation}
In order to eliminate the scalar product of two momenta, we note that
both electrons have equal average momentum; using \eqref{eq:p1} and
\eqref{eq:p2} we find \begin{equation}
\left\langle \vec{\nabla}_{12}\cdot\vec{\nabla}_{13}\right\rangle =-\frac{1}{2}\left\langle \nabla_{13}^{2}\right\rangle ,\label{eq:nablas}\end{equation}
in agreement with \cite{Frolov2007}, in whose Table II the values
of $\left\langle \vec{p}_{1}\cdot\vec{p}_{2}\right\rangle $ and $\left\langle \vec{p}_{1}\cdot\vec{p}_{3}\right\rangle $
should have minus signs \cite{FrolovPrivate}. Also equal are the
average potential energy of each electron interacting with the positron,\begin{equation}
\left\langle \frac{1}{r_{23}}\right\rangle =\left\langle \frac{1}{r_{13}}\right\rangle .\label{eq:pots}\end{equation}
Substituting the last two equalities into \eqref{eq:id}, we obtain
the main result\begin{equation}
0=\left\langle \frac{3}{2}\nabla_{13}^{2}+2\frac{1}{r_{13}}-\frac{\vec{r}_{12}\cdot\vec{r}_{13}}{r_{13}^{3}}\right\rangle =\left\langle \frac{3}{2}\nabla_{13}^{2}+\frac{1}{r_{13}}+\frac{\vec{r}_{23}\cdot\vec{r}_{13}}{r_{13}^{3}}\right\rangle =-\frac{3}{2}A+B+C.\label{eq:main}\end{equation}
It turns out that the two published expressions \eqref{eq:Liang}
and \eqref{eq:Wienczek} differ by 4/27 times this combination, that
is by zero. An evaluation of $C$ using \eqref{eq:main} and the precise
values of $A$ and $B$ \cite{Frolov2007} gives a value that differs
from \eqref{eq:C} only in the last digit (it is 0 instead of 1). 

Using \eqref{eq:main} we can eliminate $C$ in terms of $A$ and
$B$. The binding correction becomes\begin{equation}
\frac{\Delta g_{\mathrm{bound}}}{\alpha^{2}}=\left(\frac{g_{\mathrm{{free}}}}{2}-\frac{11}{9}\right)A-\frac{2g_{\mathrm{{free}}}}{3}B,\label{eq:simpler}\end{equation}
where we have included the actual value of the free-positron $g$-factor,
using the result of \cite{Wienczek2014} (rather than approximating
$g_{\mathrm{{free}}}\to2$ as in \cite{Liang:2014gwa}). Taking the recently measured
$g_{\mathrm{{free}}}$ of the electron \cite{Hanneke:2008tm} and
the values of $A$ and $B$ from \cite{Frolov2007}, we confirm and
improve the result of \cite{Wienczek2014}, as well as the less precise
result \cite{Liang:2014gwa},\begin{equation}
\Delta g_{\mathrm{bound}}=-0.510\,551\,028\,187\,6(6)\alpha^{2}+\mathcal{O}\left(\alpha^{4}\right).\label{eq:dg}\end{equation}
The error in the coefficient of $\alpha^{2}$ is dominated by the
comparison of $g$-factors of free electrons and positrons, known
to be equal to better than three parts per trillion \cite{VanDyck:1987ay}.
Of course, providing that coefficient with more than four or five
decimal places is at present only of academic interest, because of
the unknown $\alpha^{4}$ effects. Nevertheless, the improvement of
precision shows the power of the hypervirial relation that eliminates
a lesser known expectation value $C$ in favor of $A$ and $B$, very
well known from the determination of the binding energy of $\mathrm{Ps}^{-}$. 

The total $g$-factor of the bound positron is (with $\alpha$ from
\cite{Bouchendira:2010es} or from a comparison of the measured $g_{\mathrm{{free}}}$
with QED \cite{Aoyama:2012wj})\begin{equation}
g_{\mathrm{Ps}^{-}}=g_{\mathrm{{free}}}+\Delta g_{\mathrm{bound}}=2.002\,292\,117\left(3\right),\label{eq:gps}\end{equation}
where the error is estimated by $\pm\alpha^{4}$ \cite{Wienczek2014}%
\footnote{In \cite{Liang:2014gwa} the final value for $g_{\mathrm{Ps}^{-}}$ contains
an inadvertent additional term $g_{\mathrm{{free}}}-2$.%
}. 

To summarize, we have demonstrated that the two recently published
formulas for the magnetic moment of $\mathrm{Ps}^{-}$ are equivalent. Eq.~\eqref{eq:main}
is but one example of hypervirial relations that can be used to check
the accuracy of variational calculations. This is a welcome development,
now that $\mathrm{Ps}^{-}$ has become more accessible to precise measurements
\cite{PsMinusKEK}.

Acknowledgement: This reasearch was supported by Science and Engineering
Research Canada (NSERC). 


\begin{thebibliography}{10}

\bibitem{PsIonSource08}
Y. Nagashima, T. Hakodate, A. Miyamoto, and K. Michishio, New J.~Phys. {\bf
  10},  123029  (2008).

\bibitem{PsIonSource12}
H. Terabe, K. Michishio, T. Tachibana, and Y. Nagashima, New J.~Phys. {\bf 14},
   015003  (2012).

\bibitem{Liang:2014gwa}
Y. Liang, P.~L. McGrath, and A. Czarnecki, New J. Phys. {\bf 16},  063045
  (2014).

\bibitem{Wienczek2014}
A. Wienczek, M. Puchalski, and K. Pachucki, Phys. Rev. A {\bf 90},  022508
  (2014).

\bibitem{hypervirial1960}
J.~O. Hirschfelder, J. Chem. Phys. {\bf 33},  1462  (1960).

\bibitem{ShabaevVirial91}
V.~M. Shabaev, J. Phys. B {\bf 24},  4479  (1991).

\bibitem{Frolov2007}
A.~M. Frolov, J.~Phys.~A {\bf 40},  6175  (2007).

\bibitem{virialAJP}
J.~H. Epstein and S.~T. Epstein, Am. J. Phys. {\bf 30},  266  (1962).

\bibitem{FrolovPrivate}
A.~M. Frolov, private communication.

\bibitem{Hanneke:2008tm}
D. Hanneke, S. Fogwell, and G. Gabrielse, Phys. Rev. Lett. {\bf 100},  120801
  (2008).

\bibitem{VanDyck:1987ay}
R.~S. {Van Dyck Jr.}, P.~B. Schwinberg, and H.~G. Dehmelt, Phys. Rev. Lett.
  {\bf 59},  26  (1987).

\bibitem{Bouchendira:2010es}
R. Bouchendira,
P. Clade,
S. Guellati-Khelifa,
F. Nez, and F. Biraben,
 Phys.~Rev.~Lett. {\bf 106},  080801  (2011).

\bibitem{Aoyama:2012wj}
T. Aoyama, M. Hayakawa, T. Kinoshita, and M. Nio, Phys.~Rev.~Lett. {\bf 109},
  111807  (2012).

\bibitem{PsMinusKEK}
Y. Nagashima,
K. Michishio, T. Tachibana, H. Terabe, and  R. Suzuki,
 J.~Phys.: Conference Series {\bf 388},
  012021  (2012).

\end{thebibliography}

\printindex{}
\end{document}